\documentclass[preprint,12pt,authoryear]{elsarticle}

\usepackage{amssymb}
\usepackage{natbib}

\usepackage{algorithm}
\usepackage{algpseudocode}
\usepackage{hyperref} 
\usepackage{soul}

\newcommand{\bfb}[1]{{\mbox{\boldmath $#1$}}}
\newcommand{\Ab}{\textbf{A}}
\newcommand{\Bb}{\textbf{B}}
\newcommand{\Cb}{\textbf{C}}
\newcommand{\Db}{\textbf{D}}
\newcommand{\ab}{\textbf{a}}

\newcommand{\Ib}{\textbf{I}}
\newcommand{\Kb}{\textbf{K}}
\newcommand{\Lb}{\textbf{L}}

\newcommand{\Wb}{\textbf{W}}

\newcommand{\Xb}{\textbf{X}}

\newcommand{\xb}{\textbf{x}}
\newcommand{\yb}{\textbf{y}}

\newcommand{\Qb}{\textbf{Q}}
\newcommand{\Pb}{\textbf{P}}
\newcommand{\Sb}{\textbf{S}}
\newcommand{\Ub}{\textbf{U}}
\newcommand{\qb}{\textbf{q}}

\newcommand{\pb}{\textbf{p}}
\newcommand{\gb}{\textbf{g}}
\newcommand{\betab}{\bfb{\beta}}

\newcommand{\oneb}{\bfb{1}}

\newtheorem{thm}{Theorem}
\newtheorem{lem}{Lemma}
\newtheorem{pf}{Proof}

\journal{Statistics and Probability Letters}

\begin{document}

\begin{frontmatter}



\title{Solving The Ordinary Least Squares in Closed Form, Without Inversion or Normalization}  


\author[1]{Vered Senderovich Madar}
\author[2]{Sandra Batista}
\affiliation[1]{ 
           city={Chapel Hill},
            state={NC},
           country={USA}}
           
\affiliation[2]{organization={Loyola Marymount University},
addressline={1 LMU Drive, MS 8145}, 
           city={Los Angeles},
            state={CA},
           country={USA}}

\begin{abstract}
By connecting the LU factorization and the Gram-Schmidt orthogonalization without any normalization, closed-forms for the coefficients of the ordinary least squares solution are presented. Each of the coefficients is expressed and computed directly as a linear combination of vectors from a single non-normalized Gram-Schmidt process. The coefficients may be computed separately or altogether using the closed-form given. 
As immediate consequences, we also obtain a closed form for the generalized inverse, a closed-form for each of the coefficients of weighted linear regression, and a simplification of the computation of the Frisch-Waugh-Lovell Theorem. 
\end{abstract}


\begin{highlights}
\item We recognize a relationship between the LU factorization and the Gram-Schmidt orthogonalization process without normalization.

\item We use the relationship between the LU factorization and Gram-Schmidt process without normalization to offer a closed-form for each regression coefficient from the ordinary least squares solution.

\item As a result we derive a closed-form for the generalized inverse matrix.

\item We expand the relationship between LU factorization and the simplified Gram-Schmidt orthogonalization for the coefficients of weighted linear regression. 

\end{highlights}




\end{frontmatter}


\section{Introduction}\label{sec:intro}
Many data problems share common characteristics with the classical statistical solution for the problem of ordinary least squares (OLS), 
where the goal is to solve the system of linear equations between the matrix of $p$ independent random variables, $\Xb  = (\xb_1|\cdots|\xb_p)$ and a single dependent random variable $\yb$
\[
    \yb =  \Xb \betab .
\]
When the Gram matrix, $\Xb^T\Xb$, is invertible, the estimated solution for $\betab$ is 
expressed in terms of a column vector
, $\hat\betab = (\hat\beta_i)_{i=1}^p$
, and computed by the inversion of the Gram matrix:
\[
     \hat \betab =  \left(\Xb^T\Xb\right)^{-1}\Xb^T\yb .
\]

We offer closed-forms for each of the OLS coefficients, $\hat \beta_i$. The closed-forms are exact and may be calculated without computing the entire vector of coefficients $\hat\betab$ and without calculating or inverting the Gram-matrix for $\Xb$.  
They arise by clarifying the mathematical relationship between the factor $\Ub$ from $\Lb\Ub$ factorization and a simplified Gram-Schmidt orthogonalization (GSO) process for $\Xb$ that avoids normalization.
They may be considered a generalization of the exact computation of Cholesky~\citep[pp. 95]{BrezinskiTournes}.

The closed-forms we obtain for the coefficients $\hat \beta_i$ are presented in Section~\ref{sec:ClosedForms} and they are useful in many ways. 
First, they make it possible to express additional closed-forms for the generalized inverse and each of the elements in the precision matrix. This is demonstrated in Section~\ref{sec:inverse}.
Second, they can be extended for the general case of weighted linear regression.
The extension for the case of weighted linear regression is presented in Section~\ref{sec:weightedReg}. 
Finally, they provide a simplification of the Frisch-Waugh-Lovell Theorem by reducing the residualization of the dependent variable, discussed in Section~\ref{sec:FWL}. 

In the following section we present the relationship between the LU factorization and Gram-Schmidt orthogonalization used for our method. We show that for the purpose of computing the solutions for the OLS, it is possible to omit the normalization step for the GSO. 

\section{The LU factorization and Gram-Schmidt Process}\label{sec:prelmLU}

Classical exact methods to solve the OLS problem employ matrix factorization such as LU or Gram-Schmidt Orthogonalization (GSO)~\cite{GolubVanl13, Higham02}. 
These methods do not apply matrix inversion directly and instead derive an iterative algorithm for the solution for the 
 normal equations: 
\begin{equation}\label{eq:NormEq}
  \Xb^T\Xb  \betab  = \Xb^T\yb.
\end{equation}
 We review each of these methods briefly now and discuss their relationships to each other.
  The relationship between the $\Ub$ matrix from the LU Factorization and GSO is expressed in Lemma~\ref{lem:LU_Q} in Section~\ref{sec:UeqQtX}. 

\subsection{The LU Factorization}\label{sec:LUfactor}
Solving for the OLS solution is often done by applying Gaussian elimination over the normal equations~(Eq.~\ref{eq:NormEq}). The generalization of this approach is the LU factorization of the Gram matrix $\Xb^T\Xb$ into an upper triangular matrix, $\Ub$, and lower triangular matrix, $\Lb$, i.e. $\Xb^T\Xb = \Lb\Ub$.
A common approach for the LU factorization is to apply Doolittle's method~\citep[Algorithm 9.2 pp. 162]{Higham02}, an algorithm for the iterative computation of $\Ub$ and $\Lb$ all together. For the specific case of a symmetric matrix, it is sufficient to compute directly the upper triangular $\Ub$ and then to obtain $\Lb$ afterwards from $\Ub^T$. More specifically, for a symmetric matrix such as the Gram matrix, it is convenient to express the factorization as $\Lb\Db\Lb^T$, in terms of the unit lower triangular matrix $\Lb$ and the diagonal matrix $\Db = diag \left(u_{11},\cdots,u_{pp}\right)$ consisting of the diagonal elements of $\Ub$. For completeness we summarize the computation of the factor $\Ub$ for the Gram matrix as follows in Algorithm~\ref{alg:LU_known}. 
\begin{algorithm}
\scriptsize
\caption{: The computation of the factor $\Ub$ for $\Xb^T\Xb$}
\label{alg:LU_known}
\begin{algorithmic}[0]
\For{$j = 1, \ldots, p$}
  \State $u_{1,j}  = 
  <\xb_1,\xb_j>$
\EndFor
\For{$i = 2, \ldots ,p$}
    \For{$j = 1, \ldots ,p$}
       \State $u_{i,j}  = 
  <\xb_i,\xb_j> - \sum_{k=1}^{i-1} u_{ki} \frac{u_{kj}}{u_{kk}}$
    \EndFor
\EndFor
\State \Return $\Ub$ as the upper triangular factor of $\Xb^T\Xb$ 
\end{algorithmic}
\end{algorithm}

\subsection{A simplified Gram-Schmidt process}\label{sec:GSO}
The Gram-Schmidt orthogonalization process (GSO) 
transforms a given set of vectors into an orthonormalized set of vectors. The GSO process is summarized briefly in a similar vein to how it is given by Courant and Hilbert~\citep[pp. 4]{CourantHilbert}.
To indicate that the vectors are orthogonalized, but not stored normalized, we call this the Simplified Gram-Schmidt Orthogonalization Process (SGSO) for the $n \times p$ data matrix $\Xb = (\xb_1|\xb_2|,\cdots,|\xb_p)$ given as Algorithm~\ref{alg:SGSO}.

\begin{algorithm}
\scriptsize
\caption{: The Simplified Gram-Schmidt Orthogonalization (SGSO)}
\label{alg:SGSO}
\begin{algorithmic}[0]
\State $\qb_1 = \xb_1$
\For{$i = 2,\ldots,p$}
  \State $\qb_i = \xb_i - \sum_{j=1}^{i-1} \frac{<\qb_j,\xb_i>}{<\qb_j,\qb_j>}\qb_j$
\EndFor
\State \Return $\Qb = (\qb_1|\cdots|\qb_p)$
\end{algorithmic}
\end{algorithm}

The result of the SGSO is the matrix $\Qb$. As noted, the division by the inner product of orthogonal vectors without using square roots (as with the norm) is used and the orthogonal vectors of $\Qb$ are stored without normalization. This change not only simplifies exposition of the closed forms but also removes use of the square root that contributes to the numerical instability of GSO. For the closed forms and algorithms given in the following sections, we also use a version of $\Qb$ where all the column vectors are divided by their norms squared and we call such a matrix $\Qb^o =\left(\frac{\qb_1}{<\qb_1,\qb_1>}\left |\cdots \right |\frac{\qb_p}{<\qb_p,\qb_p>}\right)$. The column vectors of $\Qb^o$, denoted $\{\qb^o_i\}_{i = 1}^p$, are still not normalized. 

\subsection{The connection between U,Q 
and the weighted QR  factorization}\label{sec:UeqQtX}
In the following Lemma, the LU factor $\Ub$ will be connected to the orthogonalized but non-normalized SGSO matrix $\Qb$. We will further illustrate that the factor $\Ub$ multiplied by the weight matrix $(\Qb^T\Qb)^{-1}$ and then by $\Qb$ forms the QR factorization of $\Xb$ without the need for normalization.
\begin{lem}\label{lem:LU_Q}
Let $\Xb \in \Re^{n \times p}$, and suppose that the Gram matrix $\Xb^T\Xb$ is nonsingular. Let $\Ub \in \Re^{p \times p}$ be the upper triangular matrix constructed for $\Xb^T\Xb$ by Algorithm~\ref{alg:LU_known}, and let $\Qb\in \Re^{n \times p}$ be matrix constructed by the orthogonalization process for $\Xb$ in Algorithm~\ref{alg:SGSO}. 
Then, the upper triangular matrix $\Ub$ is connected to the $\Qb$ matrix by the relation
\begin{equation}\label{eq:U_eq_QTX}
           \Ub  = \Qb^T\Xb .
    \end{equation}
Moreover, $\Db  \equiv \Qb^T\Qb$ is a diagonal matrix consisting of the elements $u_{ii}$,      
\begin{equation}\label{eq:D}
    <\qb_i,\qb_i>   = <\qb_i,\xb_i> =  u_{ii} , \quad  \mbox{ for all } i = 1,\cdots,p
     \end{equation}
In addition, there exists an equivalent weighted QR factorization for the matrix $\Xb$ in terms of multiplication of $\Qb$, the inverse of $\Qb^T\Qb$ and $\Ub$:
\begin{equation}\label{eq:QR}
           \Xb  = \Qb\left(\Qb^T\Qb\right)^{-1}\Ub .
    \end{equation}
   Under the SGSO and LU factorization, Eq.~(\ref{eq:U_eq_QTX}) and Eq.~(\ref{eq:QR}) are equivalent.
\end{lem}
We will use Lemma~\ref{lem:LU_Q} to present two closed-forms for the coefficients $\hat \beta_i$: The weighted QR factorization in Eq.~(\ref{eq:QR}) will be used to reduce the normal equations into the seminormal equations~\cite[pp. 391-392]{Higham02}, forming a simplified triangular system of equations whose solution may be computed by backward iteration process over the elements of the upper triangular factor obtained by LU factorization over the Gram matrix of $(\Xb|\yb)$. The second closed-form will be in terms of the column vectors of $\Qb$ (or $\Qb^o$) and $\Xb$ by a further application of Eq.~(\ref{eq:U_eq_QTX}) over the iteration obtained by the first closed-form. 
The proof for the lemma is given in~\ref{sec:proofs}.

 \section{The Closed Form for the OLS Coefficients}\label{sec:ClosedForms}
Let $\Xb \in \Re^{n\times p}$ and let $\yb \in \Re^{1 \times n}$ and assume that the Gram  matrix, $\Xb^T\Xb$, is nonsingular and therefore invertible. 
However, if it is necessary to model the intercept, then an initial vector of $\oneb$ can be added before the data matrix, $\Xb$.

 Theorem~\ref{thm:beta_U} offers an iterative solution for the elements of the vector of $\hat\betab$ and does not necessarily require the computation of the SGSO matrix $\Qb$.
 Each coefficient, $\hat\beta_i$ is expressed in terms of the elements in the upper triangular LU factor for $\Xb^T(\Xb|\yb)$, the upper $p \times (p+1)$ submatrix of the LU factor $\Ub$ computed for Gram matrix, $(\Xb|\yb)^T(\Xb|\yb)$.

\begin{thm}\label{thm:beta_U}
 Assume that $\Xb^T\Xb$ is nonsingular and let $\Ub$ be the upper triangular factor from $LU$ factorization of the augmented matrix $\Xb^T(\Xb|\yb)$, and define $u_{i,y} \equiv u_{i,p+1}$, for $i = 1, \ldots, p$ (so that, $u_{1,y} = \xb_1^T\yb$ and $u_{i,y} =  \xb_i^T\yb - \sum_{k=1}^{i-1}\frac{ u_{k,i} u_{k,y}}{u_{k,k}}$). 
 Then 
the estimated coefficient, $\hat \beta_i$, can be computed in terms of ratios of the elements of $\Ub$ in recursive manner. 
 \begin{equation}\label{eq:beta_U}
 \hat\beta_i = 
 \left\{
 \begin{array}{ll}
 \frac{u_{i,y}}{u_{i,i}} -
 \sum_{j=i+1}^p \hat\beta_j \frac{u_{i,j}}{u_{i,i}}
 &  i =  1,\ldots , p - 1 \\
   \frac{u_{p,y}}{u_{pp}}  & i = p \\
   \end{array}\right.
 \end{equation}

\end{thm}

 Theorem~\ref{thm:beta_U} resembles versions of known back substitution algorithms such as the Row-Oriented Back Substitution algorithm~\citep[pp. 107, Algorithm 3.1.2]{GolubVanl13}.

In Theorem~\ref{thm:beta_SGSO} we offer a closed-form for the coefficients $\hat \beta_i$ utilizing the factorization of the LU factor $\Ub$ in terms of SGSO, as $\Ub = \Qb^T\Xb$:

\begin{thm}\label{thm:beta_SGSO}
Let $\Qb = (\qb_1|\cdots|\qb_p)$ and $\Qb^o = (\qb_1^o|\cdots|\qb_p^o)$ be from the matrix of vectors from the SGSO of the columns of $\Xb$, $\Qb$, where $\qb_j^o \equiv \frac{\qb_j}{<\qb_j,\qb_j>}$ (for $j = 1 ,\cdots,p$).
Under the assumptions of nonsingularity for $\Xb^T\Xb$ and the equality $\Ub = \Qb^T\Xb$, 
it is possible to write each of the coefficients, $\hat\beta_i$, in terms of a multiplication by $p-i$ projection matrices $\Ib  - \xb_k(\qb^o_k)^T$ for $k = i+1,\ldots,p$:
\begin{equation}\label{eq:beta_GSMO}
 \hat \beta_i   = \left\{ 
 \begin{array}{ll}
     (\qb^o_i)^T \left[\Ib - \xb_{i+1}(\qb^o_{i+1})^T\right] \cdots \left[\Ib - \xb_p(\qb^o_p)^T\right]\yb 
      &  i = 1, \cdots , p - 1 \\
     (\qb^o_p)^T\yb  & i = p \\
   \end{array} 
   \right.
   \end{equation}
  \end{thm}  

The proofs of Theorem~\ref{thm:beta_U} and Theorem~\ref{thm:beta_SGSO} are in~\ref{sec:proofs}.  
In~\ref{app:numexamp} we give a numerical example to illustrate the computation for the closed-forms in Theorem~\ref{thm:beta_U}, and in ~\ref{app:4varCF} we derive closed-forms for the coefficients $\hat\beta_4$ and $\hat\beta_3$, for the case of $p=4$ using Theorem~\ref{thm:beta_SGSO}.

\section{Results}
\label{sec:applications}

\subsection{A closed form for the generalized inverse and the precision matrix}\label{sec:inverse}

 The closed-forms of Theorem~\ref{thm:beta_SGSO} immediately give us a closed-form for the generalized inverse $\Xb^+ =  (\Xb^T\Xb)^{-1}\Xb^T$ expressed in terms of a single non-normalized Gram-Schmidt orthogonalization process (or SGSO).
\begin{thm}\label{thm:genInv}
Under the assumptions and notations of Theorem~\ref{thm:beta_SGSO}, we can exchange the dependent variable $\yb$ by the columns from the identity matrix $\Ib$ of size $n$ ($n = length(\yb)$), and obtain an orthogonalized row vector
\[
\xb_i^+ = (\qb^o_i)^T \left[\Ib - \xb_{i+1}(\qb^o_{i+1})^T\right] \cdots \left[\Ib - \xb_p(\qb^o_p)^T\right] \Ib
\]
where $\xb_i^+ \xb_i = 1$ and $\xb_i^+ \xb_j = 0$, for $i \ne j$.
In this way, as a direct result of Theorem~\ref{thm:beta_SGSO}, we can derive a closed-form  for the left-hand generalized inverse: 
\[
 \Xb^+ \equiv (\Xb^T\Xb)^{-1}\Xb^T
= 
 \left(\begin{array}{c}
(\qb^o_1)^T \left[ \Ib  - \xb_2(\qb^o_2)^T\right]\left[ \Ib  - \xb_3(\qb^o_3)^T\right]\cdots \left[ \Ib  - \xb_p(\qb^o_p)^T\right]   \\
\cdots
\\  (\qb^o_{p-2})^T \left[ \Ib  - \xb_{p-1}(\qb^o_{p-1})^T\right]\left[ \Ib  - \xb_p(\qb^o_p)^T\right]
\\ (\qb^o_{p-1})^T \left[\Ib  - \xb_p(\qb^o_p)^T\right]
 \\ (\qb^o_p)^T
\end{array} \right)
   \]
  \end{thm}  
  
Note, that when the matrix $\Xb$ is symmetric, the generalized inverse $\Xb^+$ coincides with the inverse matrix $\Xb^{-1}$.
The precision matrix, $\Sb$, can be computed directly as the Gram matrix for $\Xb^+$ as 
\[
 \Sb = (\Xb^T\Xb)^{-1}  =  \Xb^+\left(\Xb^+\right)^T
 \]
Theorem~\ref{thm:genInv} also provides a useful way to compute a single element of $\Sb = (s^{ij})_{i,j=1}^p$, in terms of the product of the vector rows of $\Xb^+$, as
$s^{ij} = \xb_i^+(\xb_j^+)^T$.
 This permits using single elements of the precision matrix without recomputing the entire precision matrix. 
Having such simplified closed-forms is useful for graphical models or covariance estimation.
  

\subsection{Extension for Weighted Linear Regression}\label{sec:weightedReg}
For the general case of weighted linear regression
the optimal solution for a linear regression model requires the use of a weight matrix $\Wb$ and the estimation for the vector of regression coefficients is as follows: 
\begin{equation}\label{eq:WOLSest}
     \hat \betab = \left(\Xb^T\Wb\Xb\right)^{-1}\Xb^T\Wb\yb .
\end{equation} 
In the case of nonhomogeneous variance, $\Wb$ will be a diagonal matrix. For models of variance components in genetic studies, the matrix $\Wb$ is computed using a kinship matrix, $\Kb$, as $\Wb = (\sigma_g^2\Kb + \sigma_\epsilon^2 \Ib)^{-1}$ where $\sigma_g^2$ is genetic variance and $\sigma_\epsilon^2$ is environmental variance \citep{SulMartinEskin18}. The matrix $\Wb$ may also represent a function of the information matrix in the case of generalized linear models.
It is common to factorize $\Wb$ into its square-root matrix, $\Wb^{1/2}$, and solve the OLS for $\Xb^* = \Wb^{1/2}\Xb$ and $\yb^*= \Wb^{1/2}\yb$ instead of $\Xb$ and $\yb$. In what follows we present an alternative form which does not require a square-root matrix factorization for $\Wb$, applies directly over the matrix $\Wb\Xb$, and allows $\Wb$ to be positive definite or indefinite.

\begin{thm}\label{thm:WOLS}
 	Let $\Xb = (\xb_1|\cdots|\xb_p)\in \Re^{n\times p}$, let $\Wb \in \Re^{n \times n}$ be a symmetric and nonsingular matrix, and define $\Qb(\Wb) = \left[\qb_1(\Wb)|\cdots|\qb_p(\Wb)\right]$ where each $\qb_i(\Wb)$ represents a column vector computed by the following process: \begin{equation}\label{eq:WSGSO}
	\qb_i(\Wb)  =
  \left\{
       \begin{array}{lcl}
       \Wb \xb_1 & i = 1 \\
       \Wb\xb_i - \sum_{j=1}^{i-1} \frac{<\qb_j(\Wb) ,\xb_i>}{<\qb_j(\Wb),\xb_j>}\qb_j(\Wb) 
	& 
  i = 2, \cdots , p 
  \end{array}\right.
		\end{equation}
Consider the solution 
$\hat\betab(\Wb) = \left(\Xb^T\Wb\Xb\right)^{-1}\Xb^T\Wb\yb$. Then, the following closed-form holds for the coefficients, of $\hat \beta(\Wb) = \left[\hat\beta_i(\Wb)\right]_{i=1}^p$:
  \[ 
       \hat\beta_i(\Wb) = 
       \left\{
       \begin{array}{lcl}
       \qb_i^T(\Wb)\left[\Ib - 
       \frac{\xb_{i+1} \qb_{i+1}^T(\Wb)} 
        {<\qb_p(\Wb),\xb_p>}\right] \cdots \left[\Ib 
        - \frac{\xb_p \qb_p^T(\Wb)} 
        {<\qb_p(\Wb),\xb_p>}\right]\yb, &
        i = 1, \cdots, p-1
       \\
       \qb_p^T(\Wb) & i =p
       \end{array}\right.
\] 
\end{thm} 
This extension is motivated by noticing that the weighted Gram matrix can be written as, $\Xb^T\Wb\Xb = \left\{<\xb_i\Wb,\xb_j>\right\}_{i,j=1}^p$, in terms of weighted dot products $\{<\xb_i\Wb,\xb_j>\}_{i,j=1}^p$. 
As a result, the upper triangular LU factor, $\Ub(\Wb)$, can be computed iteratively as in Algorithm~\ref{alg:LU_known} in terms of $<\xb_j\Wb,\xb_i>$ and $u_{ij}(\Wb)$. 
We use the connection between the factor $U(\Wb)$ and the SGSO to specify an equivalent form for the weighted SGSO matrix, denoted as $\Qb(\Wb)$. 
We finalize this section by giving a direct algorithm for computing 
 $\Xb^+ = \left(\Xb^T\Wb\Xb\right)^{-1}\Xb^T\Wb$,
which applies directly over the matrix $\Wb\Xb$.
\begin{algorithm}
\scriptsize
\caption{: Algorithm for the Generalized Inverse $\left(\Xb^T\Wb\Xb\right)^{-1}\Xb^T\Wb$}
\label{alg:WSGSO}
\begin{algorithmic}[0]
\State $\Bb = \Wb \Xb$ 
\State $d[1] = <\xb_1,\xb_1>$
\For{$i = 2,\ldots,p$}
  \State $\Bb[,i:p] = \Bb[,i:p] - B[,i-1]B[,i-1]^T  X[,i:p]/d[i-1]$
  \State $d[i] = <B[,i],\xb_i>$
\State $B[,1:i-1]  = B[,1:i-1] - B[,i]\xb_i^T  B[,1:i-1]/d[i]$
\EndFor
\State \Return $\Xb^+ =  diag(1/d)\Bb^T$
\end{algorithmic}
\end{algorithm}

 Algorithm~\ref{alg:WSGSO} is similar to algorithms for computing the GSO or SGSO. For the case of OLS or simple linear regression, one may take $\Wb$ to be the identity matrix, $\Wb = \Ib$.   
 
\subsection{A confirmation and a simplification of Frisch-Waugh-Lovell Theorem}\label{sec:FWL}

 The new closed-form in Theorem~\ref{thm:beta_SGSO} offers an alternative  confirmation for the Frisch-Waugh-Lovell (FWL) Theorem~\citep{FrischWaugh1933,Lovell1963}. As the FWL Theorem directs the computation of $\hat \beta_k$ to $\hat\beta_p$, for $k$ ($2 \le k \le p$), by a simpler linear equation applied over the residualized version of $\yb$ and $\xb_k,\cdots,\xb_p$, when the residualization is by $\xb_1,\cdots,\xb_{k-1}$. 
 Eq.~(\ref{eq:beta_GSMO}) applies directly over $\{\hat \beta_i\}_{ i < k}$, and provides the residualization  without considering the other coefficients for $\{\hat \beta_i\}_{i < k}$. 
Moreover, Eq.~(\ref{eq:beta_GSMO}) provides a simplification over FWL Theorem, as it applies directly over the dependent variable $\yb$ and does not require its additional residualization (or orthogonalization). 
 Such a relaxation may be useful, for example, for permutation tests when the permutations are applied directly over the values of $\yb$. The example of this application applied to epistasis is given in~\ref{app:epistasis}. This is demonstrated over an extensive study of pairwise epistasis for the phenotype of body mass index on a rat data set and mouse data set demonstrates this result~\cite{Batista2023}

\section{Acknowledgements}
The authors declare that they have no known competing financial interests or personal relationships that could have appeared to influence the work reported in this paper. The authors gratefully acknowledge support from the NIH under Grants R01 LM010098 and 01 AG066833R.

\appendix
\section{Examples for Theorem~\ref{thm:beta_U} and Theorem\ref{thm:beta_SGSO}}
\subsection{Numerical example for Theorem 1}\label{app:numexamp}
We illustrate the forms from Theorem~\ref{thm:beta_U} using 
the example of kidney data~\citep[pp. 4]{EfronHastie2016} where the dependent variable is a composite measure of kidney fitness (tot) modeled in a form of polynomial regression with an intercept:
\[
    tot_j \sim \beta_0 + \beta_1 age_j + \beta_2 age_j^2 , \quad j = 1,\ldots,157
\]
We compose a data matrix, $\Xb = (\oneb| age| age^2|tot)$, consisting of $4$ columns and 
$n=157$ samples. 
We compute the $4 \times 4$ Gram matrix, $\Xb^T\Xb$, and decompose it into the upper triangular matrix from the LU factorization: 
\[
\Ub = \left(\begin{array}{cccc}  
 157 &  5714 & 247514 &  ~0\\
0 & 39553.516 & 3668218 & -3108.943 \\
0 &  0 & 9674572 & -1473.118 \\
0 &  0 & 0  & 502.535\\
\end{array}\right).
\]
From $\Ub$ we compute the first 3 rows of the matrix of $\Cb = \Ub/diag(\Ub)$: 
\[
\Cb[1:3,] = \left(\begin{array}{cccc}  
 1 & 36.395 & 1576.522 &  ~0\\ 
0 & 1  &  92.7406 & -.0786\\
0 &  0 & 1 &  -.00015\\
\end{array}\right)
\]
The the upper-right $3\times 3$ submatrix of $\Cb$ provides the values for computing the regression coefficients, $\hat\beta_0,\hat\beta_1$ and $\hat\beta_2$.
We start with $\hat \beta_2$, which is computed directly from the fourth value in the third row: 
\[
\hat \beta_2 = c_{34} = -.00015
\]
To compute $\hat \beta_1$, we take $c_{24}$ and subtract the product of $\hat \beta_2$ and $c_{23}$:
\[
\hat \beta_1 = c_{24} - \hat \beta_2 c_{23} =  -.0645
\]
and the intercept, observing that the first row of $\Cb$ is the column means of $\Xb$, is then
\[
\hat \beta_0 = \bar{tot}  - \hat \beta_1 \bar{age}  - \hat \beta_2 \bar{age^2}  =     2.59.
\]

\subsection{Closed-forms for the case of 4 variables}\label{app:4varCF}
  
 Consider the estimation of the linear regression model: 
\[
   \yb = \beta_1 \xb_1 + \beta_2\xb_2+\beta_3 \xb_3+ \beta_4\xb_4,
\]
$\xb_1$ may be the constant vector of unit values (all $1$'s) or all vectors $\xb_i$ were mean centered by taking out their means.
Suppose we are interested in the estimation of 
$\beta_4$ and $\beta_3$. 
We exercise the conventional solution and use the result of the Theorem~\ref{thm:beta_SGSO}. 
For this case the estimators will be $\hat \beta_4 = \frac{<\qb_4,\yb>}{<\qb_4,\xb_4>}$ 
and $\hat \beta_3 = \frac{\qb_3^T}{<\qb_3,\xb_3>}\left[\Ib -  \frac{\xb_4 \qb_4^T}{<\qb_4,\xb_4>}\right]\yb 
$. 
We write $\qb_3 = \Pb_{.12} \xb_3$ and $\qb_4 = \Pb_{.12}\left(\Ib - \frac{\xb_3\xb_3^T\Pb_{.12}}{\xb_3^T\Pb_{.12}\xb_3}\right)
\xb_4$, using the projection matrix 
\[\Pb_{.12} \equiv \Ib - \frac{
 \xb_1\xb_1^T }{
<\xb_1,\xb_1>}
- \frac{
\left(
\xb_2 - \frac{<\xb_1,\xb_2>}{<\xb_1,\xb_1> } \xb_1\right)
\left( \xb_2 - \frac{<\xb_1,\xb_2>\xb_1}{<\xb_1,\xb_1>}  \right)^T }
{
<\xb_2,\xb_2> - \frac{<\xb_1,\xb_2>^2}{<\xb_1,\xb_1>} }\]

Finally we get, 
\[
\hat\beta_3 = 
\frac{\xb_4^T\Pb_{.12}\xb_4
\cdot \xb_3 ^T\Pb_{.12}\yb 
- \xb_3^T\Pb_{.12}\xb_4 \cdot\xb_4^T\Pb_{.12}\yb
}{
\xb_3^T\Pb_{.12}\xb_3\cdot \xb_4^T\Pb_{.12}\xb_4
- \left(\xb_3^T\Pb_{.12}\xb_4\right)^2},
\]
and
\[
\hat\beta_4 
 = 
\frac{\xb_3^T\Pb_{.12}\xb_3 \cdot \xb_4^T\Pb_{.12}\yb -  \xb_3^T\Pb_{.12}\xb_4\cdot \xb_3^T\Pb_{.12}\yb
}{
\xb_3^T\Pb_{.12}\xb_3 \cdot \xb_4^T\Pb_{.12}\xb_4
- \left(\xb_3^T\Pb_{.12}\xb_4\right)^2}.
\]

\subsection{Application for Genomic Pairwise Epistasis Detection}\label{app:epistasis}
For pairwise epistasis detection
, we investigate whether there is statistical evidence for the interaction of the pair of loci for the phenotype. To do so we consider the relationship between the quantitative phenotype, $\pb$, two genomic loci, $\gb_i$ and $\gb_j$, and their interaction $\Ib_{ij}$. This is often modeled using the following linear model:
\[
  \pb = \beta_0 + \beta_1\gb_i+\beta_2 \gb_j+ \beta_3\Ib_{ij}, 
\]
For epistasis detection we are only concerned with the magnitude of the coefficient for the interaction term, $\beta_3$, and its test statistic. 
As with the forward iterative estimate algorithm, 
We can mean center all the variables of the model to construct the matrix, $\Xb = \left(\gb_i|\gb_j|\Ib_{ij}|\pb\right)$, and apply SGSO to $\Xb$ to get the orthogonalized vector matrix $\Qb = \left(\qb_{g_i}|\qb_{g_j}|\qb_{\Ib_{ij}}|\qb_{\pb}\right)$. We can derive immediately the regression coefficient for $\beta_3$ and the corresponding $t$-test statistic as
\[
\hat{\beta_3} = \frac{ <\qb_{\Ib_{ij}}, \pb>}{<\qb_{\Ib_{ij}},\Ib_{ij}>}
\quad \mbox{ and } \quad  T(\beta_3) = \hat \beta_3 \cdot  \sqrt{(m-4)\frac{<\qb_{\Ib_{ij}}, \Ib_{ij}>}{<\qb_{\pb},\pb>}}.
\]

\section{Proofs for Lemma~\ref{lem:LU_Q}, and the Theorems}\label{sec:proofs}
\begin{pf}[Lemma~\ref{lem:LU_Q}] 
 To show that $<\qb_i,\qb_i> = <\qb_i,\xb_i>$, define $\Pb \equiv\Ib - \sum_{j=1}^{i-1} \frac{\qb_j\qb_j^T}{<\qb_j,\qb_j>}$. $\Pb$ is a projection matrix, satisfying $\Pb\Pb = \Pb$. By SGSO, $\qb_i = \Pb\xb_i$, so $<\qb_i,\xb_i> = <\Pb\xb_i,\xb_i> = <\Pb\xb_i,\Pb\xb_i> = <\qb_i,\qb_i>$.   

$<\qb_i, \xb_j> = 0$ for $i > j$. 

To show $\Ub = \Qb^T\Xb$ by mathematical induction over the rows of $\Ub$ (using the columns of $\Qb$). Since $\qb_1 \equiv \xb_1$, 
for the first row in $\Ub$, 
 $u_{1j} = \xb_1^T\xb_j =\qb_1^T\xb_j$, and specifically that $u_{11} = <\qb_1, \xb_1>$. 
For the second row of $\Ub$,
\[ 
u_{2j} = \xb_2^T\xb_j - \frac{u_{12}u_{1j}}{u_{11}}
 = \xb_2^T\xb_j  - \frac{\qb_1^T\xb_2 \cdot \qb_1^T\xb_j}{\qb_1^T\qb_1}
 = \left(\xb_2^T  - \frac{<\qb_1,\xb_2>} {<\qb_1,\qb_1>} \xb_1^T\right)\xb_j = \qb_2^T\xb_j \quad \mbox{ for } j = 1,\ldots,p
\]
Assume $u_{kj}= \qb_{k}^T\xb_j$ for all rows $1 \leq k \leq i-1$. Now for the general row $i$: 
\[
  u_{ij} = <\xb_i,\xb_j> - \sum_{k=1}^{i-1} \frac{u_{kj}u_{ki} }{u_{kk}} , \qquad j = 1 ,\ldots , p
\]
The rest follows since $u_{kk}  =  <\qb_k,\qb_k>$ and $u_{kj}  = <\qb_k,\xb_j>$ for $k < i$:
\[
   u_{ij} = <\xb_i,\xb_j> - \sum_{k=1}^{i-1} \frac{<\qb_k,\xb_j> <\qb_k,\xb_i>}{<\qb_k,\qb_k>}
   = \left(\xb_i- \sum_{k=1}^{i-1} \frac{ <\qb_k,\xb_i>}{<\qb_k,\qb_k>}\qb_k \right)^T\xb_j = \qb_i^T\xb_j.
\]

To show that $\Xb = \Qb(\Qb^T\Qb)^{-1} \Ub$, consider $\Ab \equiv \Qb(\Qb^T\Qb)^{-1} \Ub$ and express $\Qb(\Qb^T\Qb)^{-1}$ as $\Qb(\Qb^T\Qb)^{-1} = \left(\frac{\qb_1}{<\qb_1,\qb_1>}|\frac{\qb_2}{<\qb_2,\qb_2>}|\cdots |\frac{\qb_p}{<\qb_p,\qb_p>}\right)$.
 Next, for any $i$, $1 \le i \le p$ write the $i$th column vector in $\Ab$ in terms of a linear combination of the column vectors $\frac{\qb_k}{<\qb_i,\qb_i>}$ and the coefficients $u_{ik}$'s, as: 
 \[
       \ab_i = \sum_{k=1}^p \frac{u_{ik}}{<\qb_k,\qb_k>} \cdot \qb_k 
        =  \sum_{k=1}^{i-1} \frac{u_{ik}}{<\qb_k,\qb_k>} \cdot \qb_k
        + \frac{u_{ii}}{<\qb_i,\qb_i>} \cdot \qb_i 
        + \sum_{k=i+1}^p \frac{u_{ik}}{<\qb_k,\qb_k>} \cdot \qb_k 
 \]
The rest follows directly from the identity $\Ub = \Qb^T\Xb$
, with $u_{ii} = <\qb_i,\qb_i>$, $u_{ik} = <\qb_i,\xb_k>$ in general and in particular, $u_{ik} = 0$ for all $k > i$
 \[
       \ab_i  
        =  \sum_{k=1}^{i-1} \frac{<\qb_i,\xb_k>}{<\qb_k,\qb_k>} \cdot \qb_k
        + \qb_i   = \xb_i
 \]
The equality of $\ab_i =\xb_i$ follows from the formula for $\qb_i$ 
in Algorithm~\ref{alg:SGSO}.

We have proved that Eq.~(\ref{eq:QR}) follows from Eq.~(\ref{eq:U_eq_QTX}).
To show the opposite direction, that Eq.~(\ref{eq:QR}) follows from Eq.~(\ref{eq:U_eq_QTX}), consider $\Xb = \Qb\left(\Qb^T\Qb\right)^{-1}\Ub$ and  note that Eq.~(\ref{eq:U_eq_QTX})  follows straightforward  by a left-side multiplication with $\Qb^T$, and $\Qb^T\Xb = \Qb^T\Qb\left(\Qb^T\Qb\right)^{-1}\Ub = \Ub$. 
\end{pf}

 \begin{pf}[Theorem~\ref{thm:beta_U}] We may apply the weighted QR factorization, Eq.~(\ref{eq:QR}) in Lemma~\ref{lem:LU_Q} over $\Xb$ and $\Xb^T$ transforming the normal equations 
 $\Xb^T\Xb \betab = \Xb^T\yb$, 
  into the seminormal equations~\cite[pp. 391-392]{Higham02}: 
\[
   \Db^{-1}\Ub \hat\betab =  \Db^{-1}\Qb^T\yb 
   \quad \mbox{ where} \quad \Db^{-1} = {diag(\Ub)}^{-1}  =(u_{11}^{-1}, \cdots, u_{pp}^{-1})\Ib
 \]
 With the unit upper triangular matrix $\Db^{-1}\Ub$ on the left side, the seminormal equations offer a known backward iterative computation for the $\beta_i$'s.
 Moreover, by the equality between $\Db^{-1}\Ub$ and $\Db^{-1}\Qb^T\xb$, it is sufficient to consider the augmented matrix $(\Xb|\yb)$ and view $\Qb^T(\Xb|\yb)$ as the upper $p \times (p+1)$ submatrix inside the upper triangular LU factor of $(\Xb|\yb)^T(\Xb|\yb)$. 
 \end{pf}
\begin{pf} [Theorem~\ref{thm:beta_SGSO}]
Eq.~(\ref{eq:beta_GSMO}) follows from applying Lemma~\ref{lem:LU_Q} to Eq.~(\ref{eq:beta_U}) of Theorem~\ref{thm:beta_U} and factoring out the SGSO vector for the corresponding variable (i.e. $(\qb_i^o)^T$ for $\beta_i$). Start with $\hat\beta_p =  (\qb_p^o)^T\yb$. 
Write $\hat\beta_{p-1}$, explicitly as $\hat\beta_{p-1} = \left(\qb_{p-1}^o\right)^T\left(\yb -   \xb_p \hat\beta_p\right)$ then substitute the closed form for $\hat\beta_p$ to get 
\[
\hat\beta_{p-1} = \left(\qb_{p-1}^o\right)^T\left[\yb -  \xb_p (\qb_p^o)^T\yb\right] = \left(\qb_{p-1}^o\right)^T\left[\Ib - \xb_p(\qb_p^o)^T\right]\yb.
\]

To obtain the closed form for $\hat\beta_{p-2}$, using Eq.~(\ref{eq:beta_U}) and back substitution,
\[
\hat\beta_{p-2} = (\qb_{p-2}^o)^T\yb - (\qb_{p-2}^o)^T\xb_p(\qb_{p}^o)^T\yb - (\qb_{p-2}^o)^T\xb_{p-1}(\qb_{p-1}^o)^T[\Ib-\xb_p(\qb_p^o)^T]\yb
\]
\[
= (\qb_{p-2}^o)^T[\Ib -\xb_{p-1}(\qb_{p-1}^o)^T]\left[\Ib- \xb_{p}(\qb_{p}^o)^T\right]\yb. 
\]
The proof for $\hat \beta_{p-3}$
and then the proofs for general $\hat\beta_i$, $i = p - 3, \ldots,1$ will follow in a similar manner. 
\end{pf}
 \begin{pf}[Proof of Theorem~\ref{thm:WOLS}]
The proof is by noticing the following properties: First, that the matrix $\Db \equiv \Qb(\Wb)^T\Wb^{-1}\Qb(\Wb)$ is a diagonal matrix and $\left<\qb_i(\Wb),\xb_i\right> = \left<\qb_i(\Wb),\Wb^{-1}\qb_i(\Wb)\right>$ and $\left<\qb_i(\Wb),\xb_j\right> = 0$ for $j < i$. So  
	  $\Ub(\Wb) = \Qb(\Wb)^T\Xb$ is an upper triangular matrix, 
   and third, by expressing $\Xb$ as $
	\Xb =\Qb(\Wb) \Db^{-1} \Ub(\Wb)$.
 The rest of the proof follows in a straightforward manner by 
 the steps given before for the proofs 
 above.
\end{pf} 



\bibliographystyle{plainnat}
\bibliography{bibols}
\end{document}